\documentclass[aps,prb,twocolumn]{revtex4}
\usepackage{graphicx}
\usepackage{bm}
\usepackage{amsmath}
\usepackage{dcolumn}

\begin{document}

\title{Excitations in photoactive molecules from quantum Monte Carlo}
\author{Friedemann Schautz$^1$, Francesco Buda$^2$, and Claudia Filippi$^1$}
\affiliation{
$^1$Instituut-Lorentz, Universiteit Leiden, Niels Bohrweg 2, 2333 CA Leiden, 
The Netherlands\\
$^2$Leiden Institute of Chemistry, Gorlaeus Laboratoria, P.O.\ Box 9502,
2300 RA Leiden, The Netherlands}
\date{\today}

\begin{abstract}

Despite significant advances in electronic structure methods for the 
treatment of excited states, attaining an accurate description of the photoinduced
processes in photoactive biomolecules is proving very difficult. 
For the prototypical photosensitive molecules, formaldimine, formaldehyde and a minimal
protonated Schiff base model of the retinal chromophore, we investigate the performance 
of various approaches generally considered promising for the computation of excited 
potential energy surfaces.  We show that quantum Monte Carlo can accurately estimate
the excitation energies of the studied systems if one constructs carefully the trial 
wave function, including in most cases the reoptimization of its determinantal part 
within quantum Monte Carlo.
While time-dependent density functional theory and quantum Monte Carlo are generally 
in reasonable agreement, they yield a qualitatively different description of the 
isomerization of the Schiff base model.
Finally, we find that the restricted open shell Kohn-Sham method 
is at variance with quantum Monte Carlo in estimating the lowest-singlet 
excited state potential energy surface for low-symmetry molecular structures. 
\end{abstract}
\maketitle

\section{Introduction}

The absorption of visible light and its conversion to other forms of 
energy is at the heart of some of the most fundamental processes in biology.
A familiar example of light absorption initiating a biological response is the 
primary event of vision: light induces a conformational change in rhodopsin, 
the photoreceptor in the retina, which is followed by a cascade of
chemical reactions culminating in the stimulation of the optical nerve.
A microscopic understanding of light induced conformational changes 
in photoactive biomolecules is both important from a fundamental point of view and
because of existing and potential applications in biology and biotechnology.

The advances in understanding biological photosystems are so far mainly due to 
experimental discoveries since theoretical studies are currently hindered by the 
lack of a theoretical approach which is applicable to realistically large systems 
while possessing a sufficient degree of reliability.
On the one hand, several accurate quantum chemical approaches have been developed for a proper 
description of excited states but they are only applicable to relatively small systems. 
For instance, complete active space second-order perturbation theory (CASPT2)~\cite{caspt2} has been 
employed to investigate the photoisomerization mechanism in simple models of the retinal 
chromophore of rhodopsin~\cite{garavelli97,olivucci2000,olivucci2003}.  The approach is able to accurately 
describe the excited state potential energy surface along the photoisomerization path,
but it is limited to relatively small model compounds and a proper description of the important
ligand-protein interactions is still computationally prohibitive.
On the other hand, density functional theory (DFT) based approaches have a much more favorable
scaling with system size than CASPT2 and can therefore be applied to considerably larger 
molecules. In particular, the restricted open-shell Kohn-Sham method  
(ROKS)~\cite{frank98,filatov} has been recently developed to study the 
dynamics in low-spin excited states and used to model the full retinal chromophore,
including relevant parts of the protein environment~\cite{molteni}. The resulting excited 
state potential energy surface along the isomerization coordinate is qualitatively 
different from the one derived with the CASPT2 method~\cite{olivucci2000}, though the model 
systems used in these two works are different and therefore a direct comparison is not 
possible. Therefore, while the ROKS method is appealing for the low computational cost
and for the possibility of performing molecular dynamics in the excited state, its 
adequateness needs to be further validated.
Alternatively,
linear response calculations within time-dependent density functional theory 
(TDDFT)~\cite{tddft} often yield accurate excitation energies but fail for instance in
describing extended conjugated systems~\cite{c_chains} or proton transfer~\cite{dreuw} 
in excited states, that is, systems closely related 
to photoactive molecules. The capabilities and limitations of TDDFT in describing excited 
state potential surfaces of conjugated organic molecules have been extensively 
investigated in Ref.~\onlinecite{wanko2004}.

Quantum Monte Carlo (QMC) is
an alternative to conventional quantum chemical and density functional methods, and
has been successfully employed to compute ground state properties of large molecules and 
solids~\cite{qmc}.
Compared to other theoretical approaches, QMC has the advantage that it can be applied 
to sufficiently large systems and still provide an accurate description of both dynamical 
and static electronic correlation. Despite the successful use of QMC for ground state 
problems, there is relatively little experience on its application to excited 
states~\cite{grossman01,needs01,excited_nano_qmc,porphyrin}. 
The recent QMC computation of excitation energies 
of large silicon nanostructures~\cite{excited_nano_qmc} is very encouraging but 
the simple HOMO-LUMO wave functions employed there are not likely to be adequate for 
photoactive systems due to the more complex nature of their
electronic excitation.

To compare the accuracy of ROKS, TDDFT and QMC in the study of photochemical processes, 
we compute the excitation to the lowest singlet state for a set of prototypical photoactive 
molecules: formaldimine (CH$_2$NH), formaldehyde (CH$_2$O) 
and a minimal protonated Schiff base model (C$_5$H$_6$NH$_2^+$) of the retinal chromophore.
For formaldimine and the protonated Schiff base model, we find that ROKS differs 
quantitatively and qualitatively from the other methods under consideration 
at low-symmetry molecular structures.
While TDDFT excitation energies are fairly accurate in most situations, this method 
gives a qualitatively different result along a complete-active-space 
self-consistent-field (CASSCF) minimum energy path for the isomerization 
of the protonated Schiff base model.
Finally, we find that QMC provides a reliable estimate of the lowest singlet 
excitation energies of the studied molecules, provided one makes an adequate and 
careful choice of the trial wave function.  
Although simple mean-field HOMO-LUMO Jastrow-Slater wave functions 
are not always adequate for these systems, we can recover 
accurate excitations energies by using a relatively small expansion in Slater 
determinants, whose orbitals and/or coefficients are reoptimized within QMC.
    
In Sec.~\ref{methods}, we review the theoretical approaches employed in this work.
The computational details are given in Sec.~\ref{comp_details} and the numerical results
are shown in Sec.~\ref{formald} and~\ref{psb}. Finally, in Sec.~\ref{wfqmc}, we discuss the 
sensitivity of the QMC results to the choice of the trial wave function.

\section{Theoretical methods}
\label{methods}

We briefly review the theoretical methods used in this work for the computation 
of excited states, and refer for more details to the literature.

The restricted open-shell Kohn-Sham (ROKS) me\-thod~\cite{frank98,filatov} is a 
recent modification of the $\Delta$SCF approach used for the computation of multiplet 
splittings~\cite{VB,ZRB,GJ,D}. 
In the ROKS approach, the energies of the states given by single determinants 
are not computed in separate calculations as in $\Delta$SCF, but the linear 
combination corresponding to the desired state of pure symmetry is directly 
minimized under the constraint of orthogonality among the Kohn-Sham orbitals.
In particular, the energy of an open shell singlet is estimated as 
${\rm E}({\rm s})  = 2{\rm E}({\rm m}) - {\rm E}({\rm t})$, where ${\rm
 E}({\rm m})$ is the energy of the mixed singlet configuration, i.e. a single
determinant having the open shell orbitals occupied with electrons of opposite
spin, and  ${\rm E}({\rm t})$ the energy of the corresponding triplet
configuration. Within ROKS, the energy ${\rm E}({\rm s})$ is optimized using conventional 
ground state density functionals and a common set of orthogonal orbitals is used
for both contributions~\cite{frank_comment}. 

Both the $\Delta$SCF and ROKS approaches offer a practical recipe to the
computation of excited states but they cannot be fully justified from a 
theoretical point of view and their validity must be empirically corroborated. 
An appealing feature of ROKS is that the method can be easily combined with
ab-initio molecular dynamics and used to optimize the geometries in the excited 
state, access adiabatic excitations, and study the dynamics in the excited 
state~\cite{frank98,hutter2003,marx2002,frank2003_2}. In general, even though the 
ROKS method tends to underestimate the excitation energies in particular for 
$\pi\rightarrow \pi^*$ transitions~\cite{hutter2003,frank2003_1,doltsinis2003}, 
it was shown to give a 
good description of the optimal geometries of the lowest excited states of 
small organic molecules, especially for 
$n\rightarrow\pi^*$ transitions~\cite{frank98,hutter2003}.

Time-dependent density functional theory (TDDFT) is a different 
framework for the calculations of excited state properties which has become 
widely used in recent years~\cite{tddft}. The method can handle large systems 
and, differently from $\Delta$SCF or ROKS, is formally exact even though, in 
practice, one has to resort to approximate exchange-correlation functionals.
TDDFT has been extensively applied to the computation of vertical 
excitation energies since the calculation of forces within TDDFT is not 
straightforward and only recently a few implementation and applications 
of TDDFT to compute excited state geometries and adiabatic excitations have 
been published~\cite{wanko2004,vancaillie,turbomole,hutter_forces}.

Several quantum chemical approaches have been developed for a proper 
description of excited states. Methods such as multi-reference configuration 
interaction (MRCI) and complete active space second-order perturbation theory 
(CASPT2) rely on expanding, explicitly or implicitly, the wave function in Slater 
determinants. 
As the system size increases and the energies of the single-particle orbitals 
become closely spaced, the space of orbitals which must be included in the 
expansion to recover a significant fraction of electronic correlation grows 
enormously. Therefore, these techniques are very accurate but can only be 
applied to small systems. 
Even though CASPT2 was originally proposed as a method to compute excited 
state energies with an accuracy not better than 0.5 eV, it is now regarded as an 
approach which on average yields excitations in agreement with experiments to 
better than 0.2 eV~\cite{caspt2}. 
The method is quite sensitive to the construction of the active space which 
must include all important orbital excitations and is limited on current
computers to a maximum of about 15 active orbitals.

Quantum Monte Carlo techniques~\cite{qmc} is an alternative 
to density functional and conventional quantum chemistry approaches.
While many studies have demonstrated the use and reliability of QMC
for the description of ground state properties of molecular and solid systems, 
relatively little experience exists concerning 
its application to low-lying excited states. Recent studies of the excited 
states of methane, ethene, and small hydrogenated Si clusters indicate that the method 
is capable of reproducing the excitation energies of accurate quantum chemistry 
calculations~\cite{grossman01,needs01,friedemann}.
The QMC approach was also recently applied to the study of the 
excitations of large silicon nano-clusters, in combination with simple trial
wave functions~\cite{excited_nano_qmc}.
QMC methods provide a stochastic solution of the 
Schr\"odinger equation: in diffusion Monte Carlo (DMC), the imaginary-time evolution 
operator $\exp(-{\cal H}\tau)$ is used to project out the ground state from a
given trial wave function~\cite{dmc}.  
To prevent the collapse to the bosonic ground state in fermionic systems, one 
works in the fixed-node approximation, that is, finds the best solution which has 
the same nodes as a given trial wave function.  The solution is variational for 
the lowest state of a given spin symmetry belonging to a one-dimensional irreducible
representation of the point group of the molecule. It is exact for any state if the 
nodes are exact. 
Therefore, if the nodal surface of the trial wave function is a good approximation 
to the excited state one, the fixed-node constraint can be used to access accurate 
excitation energies also of states which are not the lowest in their symmetry.

The trial many-body wave function employed in this paper is of the 
Slater-Jastrow form:
\begin{eqnarray*}
\Psi_{\rm T}=\sum_{n} d_n D_n^{\uparrow} D_n^{\downarrow}
\prod_{\alpha i j}J\left(r_{ij},r_{i\alpha},r_{j\alpha}\right)\,.
\end{eqnarray*}
${\rm D}^\uparrow_n$ and ${\rm D}^\downarrow_n$ are Slater determinants of
single particle orbitals for the up- and down-spin electrons, respectively, 
and the orbitals are represented using atomic Gaussian basis. The Jastrow factor 
correlates pairs of electrons {\it i} and {\it j} with each other, and with 
every nucleus $\alpha$, and different Jastrow factors are used to describe
the correlation with different types of atoms. The parameters in the Jastrow factor 
are optimized within QMC using the variance minimization method~\cite{umrigar_vmin}. 
The Jastrow factor is positive and does not alter the nodal surface of the wave 
function which is instead fixed by the determinantal part~\cite{non_loc_pseudo}. 
Particular attention must therefore be paid to the choice of the Slater component 
which is usually a linear combination of a small number of determinants.
In the context of excited states, the complete-active-space self-consistent-field 
(CASSCF) variant of the multi-configuration self-consistent-field method (MCSCF) 
is particularly useful. These wave functions include all possible excitations 
for a given set of electrons within a chosen set of orbitals.  When the excited state 
is not orthogonal to the ground state by symmetry, the determinantal component of the 
trial wave function is obtained in a state-average MCSCF approach~\cite{samcscf}, that 
is, by optimizing an average of the ground and excited state energies. Thus, the orbitals 
represent a compromise for describing both states.

Since the optimal orbitals and expansion coefficients in the 
presence of the Jastrow factor may differ from their optimal values in 
its absence, it is important to reoptimize them in the presence of the Jastrow 
component. To this end, we extended the energy fluctuation potential (EFP) 
method~\cite{efp} to simultaneously minimize the energy with respect to the 
orbitals and the expansion coefficients of a Slater-Jastrow wave function, as 
well as to handle state averaging necessary for excited states~\cite{friedemann}. 
In the absence of the Jastrow component, the method is analogous to the MCSCF 
technique for the lowest-state of a given symmetry, and to a state-average MCSCF
approach if the excited state of interest is not the lowest in its symmetry.
Once the Jastrow factor is included, the orthogonality between the ground
and excited states is only approximately preserved in the state-average
EFP approach. The approach was tested for several singlet states of ethene 
and was shown to systematically improve the starting trial wave functions,
correcting the initial excitation energies by as much as 0.5-0.6 eV and
yielding results in excellent agreement with experiments~\cite{friedemann}.

\section{Computational details}
\label{comp_details}

The ground-state DFT, and the excited-state ROKS and TDDFT calculations 
are performed with the Car-Parrinello molecular dynamics {\tt CPMD} 
code~\cite{CP,CPMD}.
We employ the BLYP generalized gradient approximation for the exchange and 
correlation functional~\cite{Becke,LYP}, the Goedecker 
pseudopotentials~\cite{Goedecker}, an energy cutoff of 70~Ry for the 
plane-wave expansion, and a box size about 5 \AA\ larger then 
the size of the molecule. In order to avoid the inherent periodicity of a 
plane-wave calculation, we use the method described in Ref.~\cite{martyna}, which
solves the Poisson equation for non-periodic boundary conditions, thus enabling the 
study of isolated molecules. 

For formaldimine, the multi-reference configuration interaction singles and 
doubles (MR-CISD) calculations and the optimization of the excited state geometry 
within the state-average CASSCF method are performed with the {\tt COLUMBUS} quantum 
chemistry program~\cite{columbus}.
Equal weights are used in the state-average CASSCF calculations for the optimization
of the geometries.
The reference space for MRCI is of 6 active electrons in 6 orbitals and the final MRCI
energetics include Davidson corrections. It must be stressed that these MRCI calculations 
were performed with a moderate basis ($(10s6p3d)/[4s3p1d]$ for carbon and
nitrogen, and $(7s3p)/[2s1p]$ for hydrogen) and could certainly be improved. However, for the 
purpose of establishing the reliability of the other theoretical approaches, we consider 
the accuracy of the MRCI energetics to be sufficient.

For the QMC calculations, we use the {\tt CHAMP} quantum Monte Carlo code~\cite{champ} and
norm-conserving {\it sp}-non-local pseudopotentials for carbon, nitrogen and oxigen, 
generated in an all-electron Hartree-Fock calculation for the atoms~\cite{shirley}.
The orbitals in the determinantal component of the wave functions are expanded in
the Gaussian basis sets (11$s$11$p$2$d$)/[4$s$4$p$2$d$] for carbon, nitrogen, and 
oxigen, and (10$s$2$p$)/[3$s$2$p$] for hydrogen. The basis sets are 
optimized at the HF level for formaldimine and formaldehyde. 
The determinantal part of the wave function, before reoptimization in QMC, is 
generated within Hartree-Fock, CASSCF or state-average CASSCF, using the quantum 
chemistry package {\tt GAMESS(US)}~\cite{GAMESS}.  Equal weights are used in the state-average 
CASSCF calculations, and in the state-average EFP optimization of the wave function.  
The Jastrow factor contains electron-electron, electron-nucleus 
and electron-electron-nucleus terms and is described in Ref.~\onlinecite{filippi}.
For reasons of efficiency, most calculations are performed omitting 
the electron-electron-nucleus terms since the excitation energies for these systems
computed with or without the three-body terms are the same within better than 0.1 eV~\cite{non_loc_pseudo}.
The diffusion Monte Carlo time-step used for these molecules is 0.075 H$^{-1}$.
Most of the QMC results presented below are obtained in diffusion Monte Carlo.
Variational Monte Carlo (VMC) is also used to compute various expectation values 
of the trial Jastrow-Slater wave function.

\section{Results}
\label{results}

\begin{figure}[htb]
\noindent
\hspace*{0.3cm}\includegraphics[width=6.0cm]{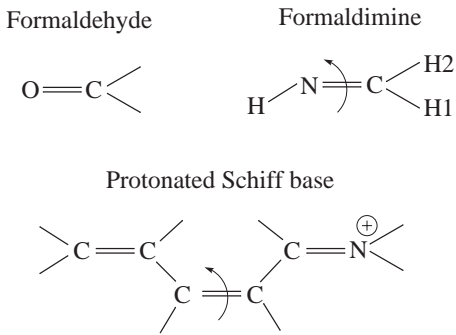}
\caption[]{Structure of the investigated molecules. In formaldimine and
the protonated Schiff base model, the isomerization is around the bond indicated
with an arrow. H1CNH is the dyhedral angle varied in formaldimine.}
\label{fig1}
\end{figure}

The photosensitive molecules we investigate are schematically shown in Fig.~\ref{fig1}.
In formaldimine and formaldehyde, the lowest singlet excitation has predominantly a 
$n\rightarrow\pi^*$ character and, in the protonated Schiff base model, a 
$\pi\rightarrow\pi^*$ character.
The performance of the DFT-based approaches may differ for the two types
of excitation, as has previously been stated for the ROKS method.

While QMC does not seem to be sensitive to the character of the excitation, a 
different complication is encountered when performing excited state QMC calculations.
If the excited state of interest is the lowest state of a given spin symmetry 
belonging to a one-dimensional irreducible representation, the DMC energy is variational. 
In all other cases, DMC is no longer variational and the quality of the trial wave 
function becomes increasingly important.
The vertical and adiabatic excitations of formaldimine and formaldehyde belong to the 
first category while the excitations of the 
minimal protonated Schiff base model and of formaldimine along its isomerization 
path belong to the other case.

\subsection{Formaldimine and formaldehyde}
\label{formald}

In the $n\rightarrow\pi^*$ excitation of formaldimine and formaldehyde, 
a lone-pair electron is transferred to a $\pi^*$ antibonding orbital. 
The excitation is almost purely of the HOMO-LUMO type and has therefore been 
considered ideal for the ROKS approach~\cite{frank98}, which was also used
to study the excited state {\it cis-trans} isomerization of formaldimine 
in a Born-Oppenheimer molecular dynamics simulation~\cite{frank98} and, 
more recently, in a non-adiabatic Car-Parrinello dynamics~\cite{marx2002}.

\begin{table}[htb]
\caption[]{Vertical and adiabatic lowest singlet excitation energies in eV for 
formaldehyde and formaldimine, calculated within ROKS, TDDFT and DMC.
The numbers in parentheses are the statistical errors on the DMC results.}
\label{table1}
\begin{tabular}{llcccc}
\hline
system    & excitation & ROKS & TDDFT & DMC     &  Expt     \\
\hline
CH$_2$O   & vertical   & 3.58 & 3.90  & 4.24(2) & 3.94$^a$, 4.07$^b$, 4.2$^c$ \\
          & adiabatic  & 3.13 & 3.51  & 3.74(2) &  3.50$^d$     \\[.5ex]
CH$_2$NH  & vertical   & 4.63 & 5.34  & 5.32(2) & 5.0--5.4$^b$  \\
          & adiabatic  & 2.85 & 3.23  & 3.21(2) & --        \\
\hline
\end{tabular}

$a$ Ref.~\onlinecite{ch2o_exp1}, $b$ Ref.~\onlinecite{ch2o_ch2nh_exp2}, $c$ Ref.~\onlinecite{ch2o_exp3},
$d$ Ref.~\onlinecite{ch2o_aexp}.
\end{table}

In Table~\ref{table1}, we list the vertical and adiabatic lowest singlet excitation energies, 
evaluated using ROKS, TDDFT and DMC. The vertical excitations are computed on the ground 
state DFT geometries, while the adiabatic excitations on the geometries optimized in the 
excited state using ROKS.  The adiabatic geometry of formaldehyde is known 
experimentally and is well reproduced by ROKS~\cite{frank98}. Vertical and adiabatic transitions are
underestimated by ROKS by as much as 0.5 eV, while the TDDFT results are in reasonable 
agreement with experiments. These findings are consistent with previous ROKS calculations
for both molecules~\cite{frank98}, 
and with TDDFT calculations of the vertical~\cite{hirata99} and 
adiabatic~\cite{turbomole} excitations of formaldehyde.

The DMC excitations are obtained using a comparable description of the ground and
excited states. A one-determinant trial wave function is used for the ground state,
and a two-determinant singlet wave function  for the excited state, corresponding to a single 
excitation from the doubly-occupied $n$ HOMO to the $\pi^*$ LUMO.
The starting orbitals in the determinantal component of the QMC wave function are from a HF 
calculation in the ground state, and a two-determinant MCSCF calculation in the excited state. 
For both states, all orbitals are subsequently optimized in the presence 
of the Jastrow factor with the EFP method.
For formaldehyde, the DMC excitation energies are slightly higher than available experimental 
numbers and results from highly-correlated quantum chemistry calculations, which however show a 
significant spread.  The vertical excitation energies computed with quantum chemistry 
techniques~\cite{gwaltney, hachey,cronstrand,vonarnim} range between 3.98~eV from EOM-CC and
4.19~eV from MRCI~\cite{vonarnim}, while MRCI calculations for the adiabatic 
transition~\cite{ch2o_lischka} yield an excitation energy of 3.60-3.66~eV. For formaldimine, the DMC 
vertical and adiabatic excitations are in good agreement with MRCI calculations~\cite{bonacic}.

While the success of DMC in describing these vertical and adiabatic excitations is
encouraging, it is important to assess its performance when variationality is lost
as happens along the low-symmetry isomerization path induced by the excitation.
We therefore consider the prototypical case of the isomerization of formaldimine around 
the C-N double bond. 
The isomerization path is constructed by constraining the torsional angle H1CNH 
(see Fig.~\ref{fig1}) at values between 0 and 90 degrees, with increments 
of 15 degrees. The molecule has $C_s$ symmetry at 0 and 90 degrees, and no symmetry 
at intermediate angles.

\begin{figure}[bht]
\noindent
\hspace*{0.3cm}\includegraphics[width=7.5cm]{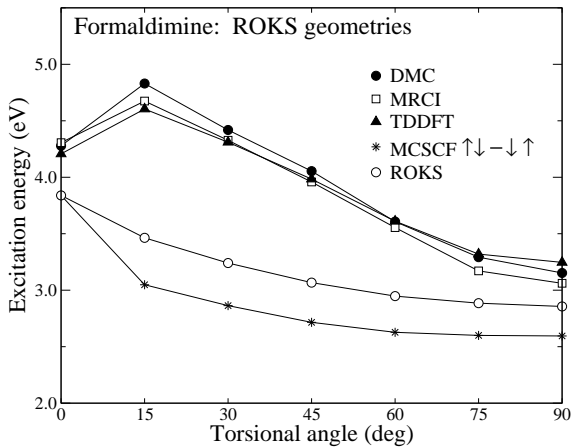}
\vspace{1.5ex}
\caption[]{Lowest-singlet excitation energies of formaldimine in eV calculated with 
ROKS, TDDFT, MRCI and DMC on the excited state geometries optimized with ROKS at constrained 
torsional angles.  The excitations computed within a two-determinant MCSCF calculation 
($\uparrow\downarrow-\downarrow\uparrow$) are also shown. The statistical error on the
DMC results is smaller than the size of the symbols.}
\label{fig2}
\end{figure}

In Fig.~\ref{fig2}, we show the ROKS, TDDFT, DMC and MRCI excitation energies 
on the excited state geometries optimized with ROKS at constrained torsional angles.  
The excitation energies are given with respect to the ground state energy
consistently computed within the same approach on the DFT ground state geometry at 
zero torsional angle.
The DMC excited state energies are obtained with a trial wave function from a 
state-average CASSCF with an active space of 6 electrons in 6 orbitals, 
whose expansion coefficients are then reoptimized in the presence of the 
Jastrow factor with a state-average EFP method. 
The DMC ground state energy at zero torsional angle is
computed with an unoptimized HF determinantal component.  The DMC excitations are in 
very good agreement with the MRCI values, with a maximum deviation of 0.13 eV 
along the curve.

While the TDDFT excitations agree with the MRCI values to 
better than 0.2 eV, the ROKS curve differs significantly.
In particular, MRCI gives a barrier to isomerization along the geometries corresponding 
to an energy minimum path in ROKS.  
One can possibly understand the behavior of ROKS by looking at the results obtained
with a two-determinant MCSCF (without state-average) approach along the
same path. 
As shown in Fig.~\ref{fig2}, the two-determinant MCSCF curve
is qualitatively very similar to the ROKS curve. 
For the two-determinant MCSCF calculation, only the orthogonality
constraint on the open shell orbitals keeps the wave function from completely
collapsing to the ground state. By analogy, the ROKS approach is likely to
suffer from the same problem whenever ground and excited states do not belong
to different irreducible representations~\cite{hutter2003}.

\begin{figure}[bht]
\noindent
\hspace*{0.3cm}\includegraphics[width=7.5cm]{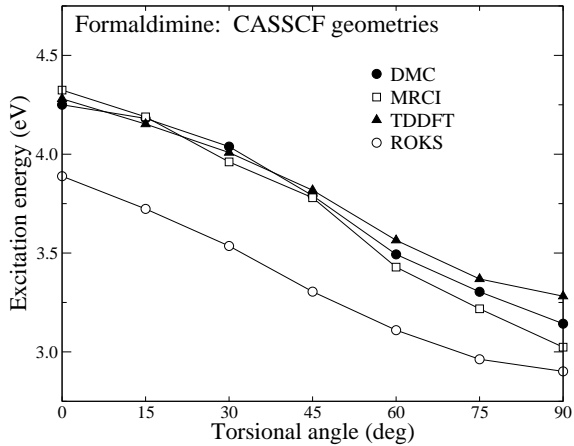}
\vspace{1.5ex}
\caption[]{Lowest-singlet excitation energies of formaldimine in eV calculated with ROKS, TDDFT, 
MRCI and DMC on the excited state geometries optimized using a state-average CASSCF 
(see text) at constrained torsional angles.}
\label{fig3}
\end{figure}

To further investigate the constraint isomerization path of formaldimine,
we optimize the geometries using the excited-state forces from
a state-average CASSCF approach with an active space of 6 electrons in 6 orbitals.
As already pointed out in early MRCI studies by Bona\v{c}i\'c-Kouteck\'y 
{\it et al.}~\cite{bonacic2}, to properly describe the isomerization of 
formaldimine, one should map the potential energy surface with respect to the 
CNH valence angle and a properly symmetrized dyhedral angle. However, the path 
obtained within CASSCF by only constraining the H1CNH dyhedral angle is reasonably 
close to the optimal path. 
We find that the main difference between the ROKS and CASSCF paths is in the behavior of the
angle CNH which, in ROKS, takes his final value corresponding to a torsional angle of 
90 degrees as soon as the molecule is displaced from planarity.

The excitations computed with TDDFT, ROKS, DMC and MRCI on the CASSCF geometries are
shown in Fig.~\ref{fig3}. 
The DMC 
calculations are performed with the same type of wave function previously used for the 
ROKS path. 
The energy barrier to isomerization present in Fig.~\ref{fig2} disappears in MRCI as 
this barrier was an artifact of using the geometries optimized within ROKS.  
The DMC excitation energies are very close to the MRCI values with a maximum 
difference of 0.1 eV along the CASSCF path.  TDDFT is in reasonable agreement with 
QMC also along this path.
For the CASSCF geometries, ROKS calculations
produce a curve of similar shape as those obtained with the 
other methods, but significantly shifted toward lower energies.

\subsection{Protonated Shiff base model}
\label{psb}

The C$_5$H$_6$NH$_2^+$ protonated Shiff base  molecule represents a 
minimal model for studying the retinal photoisomerization process in rhodopsin. 
Given its relevance and combined simplicity, this molecule is ideal for accessing 
the relative accuracy of different theoretical approaches.
Moreover, this model has been extensively studied within CASPT2 using geometries 
optimized in the excited state with CASSCF~\cite{garavelli97,olivucci2000} and, 
more recently, with CASPT2~\cite{olivucci2003}. 

\begin{figure}[bht]
\noindent
\hspace*{0.3cm}\includegraphics[width=7.5cm]{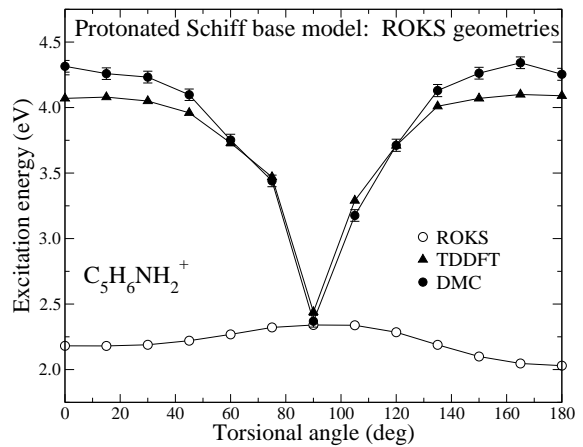}
\vspace{1.5ex}
\caption[]{Lowest-singlet excitations energies for the protonated Schiff base model in eV calculated 
with ROKS, TDDFT and DMC on the excited state geometries optimized with ROKS at constrained 
torsional angles. The excitation energies are given with respect to the ground state energy
consistently computed within the same approach on the DFT ground state cis-geometry at zero
torsional angle.}
\label{fig4}
\end{figure}

Since ROKS was previously employed to study the excited state of the full retinal 
chromophore including relevant parts of the protein environment~\cite{molteni},
it is interesting to use the same approach to optimize the structure of this simpler 
model. 
In Fig.~\ref{fig4}, we show the ROKS, TDDFT and DMC energetics computed on the 
geometries optimized within ROKS along the relevant isomerization coordinate 
represented by the torsional angle around the central C-C double bond (see Fig.~\ref{fig1}).
When optimizing the excited state geometry with ROKS, the molecule remains 
planar and the main effect of the excitation is a considerable lengthening of the 
double bonds and a shortening of the single bonds, thus reversing the conjugation 
of the molecule. 
The ROKS potential energy surface along the torsion is quite flat with a maximum at 90 degrees. 
This behavior is qualitatively different from the CASSCF and CASPT2 energy 
profile~\cite{garavelli97}, where the torsion accelerates the system towards the 
conical intersection, thus spontaneously inducing the photoisomerization. 
Therefore, while the ROKS method shows a stretching mode starting from the 
Franck-Condon region similar to the CASSCF result, it does not reproduce the 
qualitative shape of the excited state CASSCF potential energy surface along the 
torsional mode.

The DMC excited state energies in Fig.~\ref{fig4} are computed on the ROKS geometries 
with a trial wave function from a state-average CASSCF with an active space of 6 electrons 
in 6 orbitals, whose expansion coefficients are then reoptimized in the presence of the
Jastrow factor with a state-average EFP method. The TDDFT excitation energies are 
higher than the ROKS values by as much as 2 eV, and in agreement with the DMC results 
to better than 0.2 eV. The TDDFT and DMC potential energy curves have a very different 
shape than the one obtained within ROKS.
In the protonated Schiff base model, the ground and excited states belong to 
the same irreducible representation both when the molecule is planar and twisted.
The behavior of ROKS can possibly be explained as due to a contamination 
of the excited state with the ground state as in the case of twisted formaldimine.

To allow for a comparison with existing CASPT2 calculations on this model, we 
consider three geometries which were optimized in Ref.~\onlinecite{garavelli97} within 
state-average CASSCF and where the CASPT2 energies are also available. 
These structures correspond to the ground state cis-configuration 
where the Franck-Condon (FC) excitation is computed, to the geometry which demarcates 
where torsion becomes dominant along the isomerization path (denoted with HM in 
Ref.~\onlinecite{garavelli97}), and to the S$_0$/S$_1$ conical intersection (CI).
Without a direct comparison with experimental data, it is difficult to access 
the accuracy of these excited state structures: 
for instance, when compared to geometries optmized with CASPT2, the CASSCF structures
are very similar at the conical intersection but significantly
different at constrained planar symmetry~\cite{olivucci2003}.

\begin{table}[htb]
\caption[]{Lowest-singlet excitation energies for the protonated Schiff base model in eV,
calculated with TDDFT, CASPT2 and DMC on the ground state cis-configuration 
(FC), on the geometry (HM) which demarcates where torsion becomes dominant, and on 
the conical intersection (CI). The CASSCF geometries and the CASPT2 numbers are 
from Ref.~\onlinecite{garavelli97}. The excitation energies are given with respect to 
the ground state energy consistently computed within the same approach on the 
CASSCF ground state cis-geometry at zero torsional angle.}
\label{table3}
\begin{tabular}{lccc}
\hline
Geometry & TDDFT & CASPT2 & DMC \\
\hline
FC  & 3.90 &  4.02 &  4.38(5) \\
HM  & 4.12 &  3.71 &  4.22(5) \\
CI  & 2.18 &  2.19 &  2.58(5) \\
\hline
\end{tabular}
\end{table}

In Table~\ref{table3}, we list the TDDFT, CASPT2 and DMC excitation energies at 
the FC, HM and CI geometries.  The DMC calculations are performed with the same type 
of wave function previously used for the ROKS path. The use of larger active spaces
(6 electrons in 9 orbitals or 8 electrons in 8 orbitals) and the reoptimization of 
the active orbitals with the state-average EFP method yield DMC energies compatible 
to better than 0.1 eV.
While the CASPT2 and QMC results are qualitatively 
similar, the CASPT2 energies are lower than the QMC values by as much as 0.5 eV. 
The order of the TDDFT excitation energies at the FC and HM configurations are
instead reversed with respect to the DMC values: the TDDFT excitation is lower at FC
than at HM, so TDDFT gives a barrier to isomerization along the CASSCF path.
A valid question is whether this barrier survives when using an excited state path 
fully optimized within TDDFT. Recently, it has been shown that the TDDFT gradient for 
various protonated Schiff base models differs qualitatively from that of CASSCF/CASPT2, 
driving the system from the FC point to a planar fictitious stationary point~\cite{wanko2004}.

\begin{figure}[bht]
\noindent
\hspace*{0.3cm}\includegraphics[width=7.5cm]{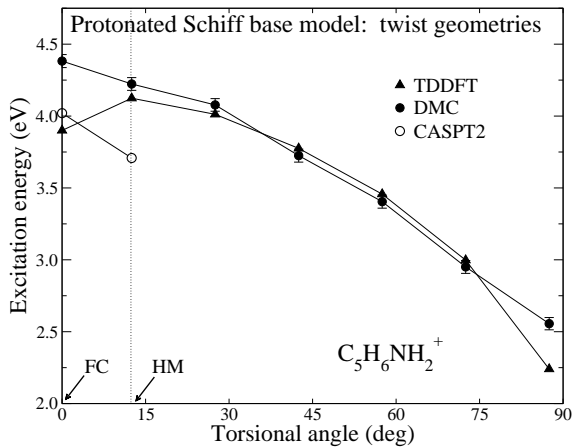}
\vspace{1.5ex}
\caption[]{Excitation energies for the protonated Schiff base model in eV,
calculated with TDDFT and DMC on a set of geometries generated by rigidly 
increasing the torsional angle, from the HM configuration. The TDDFT, DMC and
CASPT2~\cite{garavelli97} energies at FC and HM are also given.}
\label{fig5}
\end{figure}

Finally, in order to further compare TDDFT and QMC, we generate a set of geometries 
for C$_5$H$_6$NH$_2^+$ by starting from the HM structure of Ref.~\onlinecite{garavelli97} 
and increasing the torsional angle up to about 90 degrees while keeping all the other 
internal coordinates fixed. 
In Fig.~\ref{fig5}, we show the TDDFT and DMC energies, and the CASPT2 results
at FC and HM. Along the torsional path after HM, TDDFT and DMC follow closely each 
with a larger deviation at the end of the path.

\subsection{Sensitivity of DMC to the trial wave function} 
\label{wfqmc}
  
Using as examples the vertical excited state and the adiabatic isomerization path of 
formaldimine, we demonstrate how sensitive the QMC energies are to the choice of the wave 
function and how this sensitivity can vary along the excited state potential 
energy curve.

The vertical lowest-singlet excited state of formal\-di\-mi\-ne does not have a strong 
multi-configurational character, and a two-determinant Jastrow-Slater wave function to 
preserve spin symmetry is found to be sufficient for this particular state. 
The QMC energies are variational since this excited state is the lowest in its symmetry,
and orthogonality between ground and excited state is automatically ensured.
For the ground state, a single determinant wave function gives an adequate description.
In Table~\ref{table4}, we show the VMC and fixed-node DMC energies determined with different 
choices of orbitals in the determinantal component of the wave function.
The starting trial wave function uses orbitals obtained from a HF and a two-determinant 
MCSCF calculations for the ground and excited state, respectively.
By optimizing the orbitals with the EFP method, the VMC energy drops by 10 mHartree in the
ground state and by 15 mHartree in the excited state. However, the gain in the DMC energies
is only of a few mHartree and is actually more significant in the ground state. The 
resulting DMC excitation energy is only slightly higher as a result of the optimization.

\begin{table}[htb]
\caption[]{VMC and DMC ground state (S$_0$) and lowest-singlet excited state (S$_1$) 
energies in Hartree for formaldimine, calculated at the ground state geometry.
In the Jastrow-Slater wave function, a single determinant is used for the ground state
and two determinants for the excited state. The DMC excitation energies in eV are computed 
using unoptimized (HF for S$_0$ and MCSCF for S$_1$) and optimized (EFP) orbitals 
for both states.}
\label{table4}
\begin{tabular}{llccc}
\hline
State    & Orbitals & $E_{\rm VMC}$  & $E_{\rm DMC}$ & $\Delta E$ (eV) \\
\hline
S$_0$    & HF       & -17.2973(4) &  -17.3685(5)     & -- \\
         & EFP      & -17.3082(4) &  -17.3726(5)     & -- \\
\hline
S$_1$    & MCSCF    & -17.1185(4) &  -17.1756(5)     & 5.25(2) \\
         & EFP      & -17.1334(4) &  -17.1772(4)     & 5.32(2) \\
\hline
\end{tabular}
\end{table}

\begin{figure}[bht]
\noindent
\hspace*{0.3cm}\includegraphics[width=7.5cm]{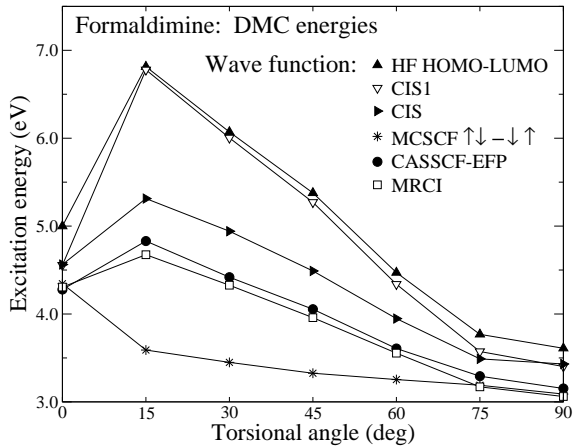}
\vspace{1.5ex}
\caption[]{DMC lowest-singlet excited state energies of formaldimine in eV, computed
on the ROKS geometries at various torsional angles, using different trial wave
functions. See text for more details.}
\label{fig6}
\end{figure}

Along the isomerization path of formaldimine, orthogonality between ground and excited 
state is no longer maintained and a higher sensitivity of the QMC results to the trial
wave function may be expected than in the case of the vertical excitation.
In Fig.~\ref{fig6}, we compare the DMC excitation energies along the ROKS 
isomerization path of formaldimine for different choices of wave functions
previously employed in other QMC studies of excited states. 
At 0 and 90$^\circ$ torsional angles where the energy is variational due to symmetry, the
spread of the DMC energies due to the use of different wave functions is 
significantly smaller than at intermediate angles.
A simple two-determinant HOMO-LUMO wave function with HF orbitals shows 
a discrepancy as large as 1.5 eV with our best DMC results obtained with a 
6 electrons in 6 orbitals CASSCF wave function whose CI coefficients have been 
reoptimized with the state-average EFP method.
The wave function denoted with CIS1 includes all single excitations from
the HOMO, and can be resummed to two determinants, where only the LUMO 
has therefore been changed with respect to the HF orbitals.
The CIS1 energies represent an improvement at the end points of the path 
but remain as poor as those obtained with a HOMO-LUMO wave function at almost
all other angles.
If all single excitations are included in a CIS wave function, 
the excitation energies are significantly closer to the CASSCF-EFP results along the 
whole path, with an almost constant discrepancy of 0.3-0.5 eV. Finally, one 
could be tempted to use a two-determinant wave function obtained in a MCSCF 
calculation (without state-average). While this wave function performs well at 0 and 90 degrees 
where ground and excited states are orthogonal by symmetry, it represents 
a poor starting point at low-symmetry configurations as already discussed
in Section~\ref{formald}, yielding DMC energies which are obviously non
variational.

Finally, the effect of
truncating the determinantal expansion according to a threshold on the coefficients is 
investigated. It is indeed customary in QMC to apply a threshold for computational 
efficiency, justified by the very different role of the reference wave function in
QMC compared to conventional quantum chemistry methods.
A smaller number of determinants is needed in a Jastrow-Slater wave function since 
the reference wave function does not define the single-particle excitation space for 
the description of dynamical correlation as is the case for a method like 
MRCI. Moreover, one hopes that the effect of determinants with a 
small coefficient on the nodal surface of the total wave function is not significant.

\begin{table}[htb]
\caption[]{VMC and DMC lowest-singlet excited state energies for formaldimine, computed
on the ROKS geometries at various torsional angles. Different determinantal components 
are used in the trial wave functions, with thresholds of 0.1 and 0.01 on the expansion 
in symmetry-adapted configuration state functions from a state-average CASSCF, and
with CASSCF and EFP-optimized expansion coefficients.}
\label{table5}
\begin{tabular}{lccc}
\hline
Threshold     & 0.1    & 0.01   & 0.01 \\
Coefficients  & CASSCF & CASSCF & EFP  \\
\hline
Angle (deg) & \multicolumn{3}{c}{Number of determinants}\\
 0  & 4   & 42    & 23  \\   
30  & 9   & 132   & 46  \\
60  & 8   & 108   & 54  \\
90  & 4   & 71    & 35  \\
\hline
Angle (deg) & \multicolumn{3}{c}{VMC energies (Hartree)}\\
 0  &  -17.158(1)  &  -17.152(1)  &  -17.165(1) \\   
30  &  -17.149(1)  &  -17.144(1)  &  -17.158(1) \\
60  &  -17.180(1)  &  -17.178(1)  &  -17.190(1) \\
90  &  -17.200(1)  &  -17.193(1)  &  -17.205(1) \\
\hline
Angle (deg) & \multicolumn{3}{c}{DMC energies (Hartree)}\\
 0  & -17.2099(5) &  -17.2063(5) &  -17.2113(4) \\
30  & -17.2027(5) &  -17.2000(5) &  -17.2062(4) \\
60  & -17.2338(5) &  -17.2313(5) &  -17.2360(4) \\
90  & -17.2502(4) &  -17.2474(5) &  -17.2527(4) \\
\hline
\end{tabular}
\end{table}

In Table~\ref{table5}, we show the VMC and DMC excited state energies for 
formaldimine, computed on the ROKS geometries at various torsional angles when
applying two different thresholds on the expansion coefficients in symmetry-adapted 
configuration state functions. 
The starting trial wave function is obtained from a state-average CASSCF
with an active space of 6 electrons in 6 orbitals.
As the threshold is lowered from 0.1 to 0.01, both VMC and DMC energies become 
higher at all angles. Since at 0 and 90 degrees the energies are variational due to
symmetry, one is unequivocally aiming at obtaining the lowest possible energy at those
geometries and one would have expected a lowering of the energy by including more
configurations. 
This indicates that the result is strongly dependent on the chosen threshold 
if one does not reoptimize the determinantal expansion in the presence of the 
Jastrow factor.  
The coefficients of the starting CASSCF wave function are therefore reoptimized with 
the state-average EFP method. The natural orbitals of the averaged single-particle
density matrix of the reoptimized expansions are here used to obtain a more compact 
wave function, and a threshold of 0.01 is then applied.
The corresponding VMC and DMC energies are also shown in Table~\ref{table5}.
At all angles, the VMC energies for the reoptimized wave function are lower 
than the values obtained using the original CASSCF coefficients with respect to 
the same threshold. Moreover, the optimal energies are also systematically better 
than the VMC values obtained with a threshold of 0.1. 
In Table~\ref{table5}, we also list the number of determinants with coefficients 
greater than the chosen threshold.  As expected, due to the inclusion of dynamical 
correlation through the Jastrow factor, the wave function becomes more compact as
an effect of the reoptimization.
The DMC energies behave similarly to the VMC values with respect to both threshold
and reoptimization. The excitation energies obtained in DMC with the reoptimized wave 
function are in excellent agreement with the MRCI values as shown in Section~\ref{formald}.
If a threshold of 0.1 is used when reoptimizing the expansion coefficients in a state-average EFP 
method, there is no improvement in the QMC energies compared to the values obtained with 
the original CASSCF coefficients and the same threshold. 

Finally, if the orbitals 
are optimized with the state-average EFP approach and a threshold of 0.1, both VMC and DMC 
energies improve and became equal to the values obtained with the CASSCF-EFP with 0.01 threshold. 
For instance, for a torsional angle of 30 degrees, the optimization of the orbitals yields a VMC 
and a DMC energy of -17.156(1) and -17.2071(4) Hartree, respectively.
We want to stress that there is in general no justification for using a threshold 
as high as 0.1 and the apparent agreement with the optimized energies is here a fortunate
case.  

\section{Conclusions}

Using TDDFT, ROKS and QMC, we have investigated the lowest-singlet excitation 
energies along various isomerization paths for the following representative 
photoactive molecules: formaldehyde, formaldimine and a minimal 
protonated Schiff base model C$_5$H$_6$NH$_2^+$.

We show that fixed-node diffusion Monte Carlo can give accurate 
excitation energies, provided a careful choice of QMC trial wave function is made. 
While simple HOMO-LUMO trial wave functions are not always adequate 
to describe the excited states of these photoactive molecules, accurate results are 
recovered when using a relatively small expansion in Slater determinants, whose 
coefficients and/or orbitals are reoptimized in the presence of the Jastrow factor 
with the EFP method.

TDDFT yields excitation energies which are generally in reasonable agreement 
with the QMC results. However, the TDDFT energies for the minimal model of the 
retinal chromophore are in qualitative disagreement with QMC and CASPT2, giving 
a barrier to isomerization along the CASSCF minimal energy path. 
 
We find that the ROKS method does not produce reliable results 
for the excited-state potential energy surface at low-symmetry configurations.  
The major source of error in the ROKS approach 
seems to be the contamination of the excited state with the ground state.
For example, ROKS predicts an energy barrier to isomerization with a maximum at 
90 degrees along the relevant torsional angle of the minimal protonated Schiff base
model of the retinal chromophore, while TDDFT and QMC show a minimum at this point.
Therefore, even though the ROKS method is appealing for its simplicity in computing 
forces, it should be generally used with caution in excited-state molecular dynamics 
simulations.

\acknowledgements
We thank J. Hutter for helpful discussions and C.\ J.\ Umrigar for a critical reading
of the manuscript.
This work is in part funded by the Stichting voor Fundamenteel Onderzoek der 
Materie (FOM), which is financially supported by the Nederlandse Organisatie voor Wetenschappelijk
Onderzoek (NWO).


\begin{thebibliography}{99}
\bibitem{non_loc_pseudo} If non-local pseudopotentials are used, the trial wave function
including the Jastrow factor is used to localize the pseudopotential so that the DMC energy 
will depend on the Jastrow factor.
\end{thebibliography}
\end{document}